\documentclass[preprint,showpacs,superscriptaddress,preprintnumbers,amsmath,amssymb,amsthm,nofootinbib]{revtex4}

\usepackage{graphicx, color}
\usepackage{dcolumn}
\usepackage{bm}

\newcommand\p{\partial}

\usepackage[normalem]{ulem}  

\newtheorem{theorem}{Theorem. }

\begin{document}

\preprint{RIKEN-QHP-182, RIKEN-STAMP-10}

\title{
Jarzynski-type equalities in gambling:
role of information \\ 
in capital growth 
}

\author{Yuji~Hirono}
\email{yuji.hirono@stonybrook.edu}
\affiliation{
Department of Physics and Astronomy, Stony Brook University,
Stony Brook, New York 11794-3800, USA
}

\author{Yoshimasa Hidaka}
\email{hidaka@riken.jp}
\affiliation{
Theoretical Research Division, Nishina Center, RIKEN, Wako 351-0198, Japan
}

\date{\today}

\begin{abstract}
We study the capital growth in gambling with (and without) side
information and memory effects. 
We derive several equalities for gambling, which are of similar form to 
the Jarzynski equality and its extension to systems with feedback controls.
Those relations provide us with new measures to quantify the effects of
information on the statistics of capital growth in gambling. 
We discuss the implications of the equalities and show that 
they reproduce the known upper bounds of average capital growth rates. 
\end{abstract}

\maketitle

\tableofcontents

\newpage

\section{Introduction}

The development of non-equilibrium equalities, such as the fluctuation
theorems and the Jarzynski equality, 
is one of the major advances in statistical physics since 1990's 
\cite{evans1993probability, 
evans1994equilibrium, 
jarzynski1997nonequilibrium, 
crooks1999entropy,
lebowitz1999gallavotti, 
kurchan2000quantum,
tasaki2000jarzynski,
hatano2001steady,
PhysRevLett.92.230602, 
harada2005equality,
PhysRevLett.95.040602,
andrieux2007fluctuation,
PhysRevLett.102.210401}.  
Those relations entail the information about the fluctuations of 
the entropy production in a stochastic environment, and are used in
deriving the second law of thermodynamics. 
They are generalized to the systems with 
measurements and feedback controls, and the relation between the
extractable work and the information obtained by the measurement is clarified
\cite{sagawa2008second, sagawa2009minimal, sagawa2010generalized};
thereby the paradox of Maxwell's demon is fully understood
\cite{sagawa2009minimal}. 
Incorporation of information processing into thermodynamics is 
a topic of considerable interest recently 
\cite{mandal2012work,barato2013autonomous,mandal2013maxwell,barato2014unifying,
deffner2013information,barato2014efficiency,parrondo2015thermodynamics,horowitz2014second}. 

Information is a key to success for gamblers, as well as for demons. 
The purpose of this paper is to argue that a similar theoretical
framework can be developed for gambling\footnote{
The analogy between the work extraction in a feedback control
system and gambling is recently discussed in Ref.~\cite{vinkler2014analogy}. 
}. 
The close connection between the capital growth rate in gambling 
and information theory is first noted by Kelly in 1956 \cite{kelly1956new}. 
He showed that the optimal growth rate of wealth in repeated horse races 
is bounded from above by an information-theoretical quantity, which is the
channel capacity \cite{shannon1949communication}. 
Kelly also considered a gamble with a help of an informant who tells 
the results (which can be wrong) to the gambler before betting. 
Kelly pointed out that the increase in the maximum of the average capital
growth rate because of the information from the informant,  
called ``side information,'' 
is quantified by the mutual information between the side information and
the results of the gamble. 
The idea to maximize the log of the capital, called the Kelly
criterion, is first applied to actual gambling 
\cite{thorp1969optimal} and stock investing \cite{thorp1971portfolio} 
by Thorp, who is the inventor of the card counting\footnote{
Card counting is a method to improve a gambler's return by utilizing the
information of dealt cards. 
} in blackjack
\cite{thorp1966beat}. 
The theory of Kelly is introduced into a wider audience by the book
``Fortune's formula'' written by Poundstone
\cite{poundstone2010fortune}. 

In this paper, we derive several equalities that 
constrain the statistics of the capital growth rate in gambling. 
We first consider a repeated gambling of binary options, each game of
which is independent and 
identical, and we derive Jarzynski-type equalities in this case. 
In the presence of side information, the equality includes 
the mutual information between the
side information and the outcome of the gambling. 
We then extend the equalities to the cases with memory effects, in which 
the outcome of a game can be dependent on the results in the past. 
In the case of gambling with side information, 
we obtain an equality involving the directed information 
\cite{massey1990causality, kramer1998directed}, 
which is a measure of causal correlations.  
The derived Jarzynski-type equalities elucidate the role of 
information in capital growth in gambling. 
We find that, applying Jensen's inequality to the equalities, 
the upper bounds of capital growth rates are reproduced, 
which is analogous to the derivation of the second law of thermodynamics
from the Jarzynski equality \cite{jarzynski1997nonequilibrium}. 
Some of the equalities in the case of independent games of gambling that we
show here are also derived by Bell and Cover using a different line of
reasoning \cite{doi:10.1287/moor.5.2.161}. 
In that sense, our main contribution lies in the generalization of those
relations to gambling which has memory effects.
Still, 
for the sake of illustration, we start with discussing 
independent games of gambling.

The paper is organized as follows. 
In Sec.~\ref{sec:kelly}, we review the Kelly criterion in games of
binary gambling which are independent and identical. 
We illustrate how the upper
bound of the average capital growth rate is related to the concepts in 
information theory. 
In Sec.~\ref{sec:jarzynski}, we derive several Jarzynski-type
equalities for independent games of binary gambling, and discuss their
implications. 
We introduce a quantity called efficacy as a measure of how well the
gambler bets as well as the quality of side information. 
In Sec.~\ref{sec:jarzynski-mem}, we extend the Jarzynski-type equalities
for the binary gambling with memory effects, in which the results can be
dependent on the ones in the past. 
We further generalize the relation to horse races, which includes the
binary betting as a special case. 
The efficacy is also generalized to the case with memory effects. 
In the last section of Sec.~\ref{sec:jarzynski-mem}, we give the unified 
expression of the Jarzynski-type equalities in gambling discussed in
this paper. 
Section \ref{sec:summary} is devoted to summary and outlook. 
In Appendix \ref{sec:app-notations}, we summarize the notations for
information-theoretical quantities used in this paper.  
In Appendix \ref{sec:app-ising}, we discuss how the Markovian coin
tossing 
discussed in Sec.~\ref{sec:jarzynski-mem} is mapped to the 1D
Ising model.

\section{Kelly criterion and information theory}\label{sec:kelly}

Let us start by giving a pedagogical review on the Kelly criterion and its
relation to information theory, using a repeated binary gambling 
where the results are independent and identically distributed ({\rm i.i.d.}). 
We demonstrate that the upper bound of the
average capital growth rate is written in terms of quantities in
information theory. 
In Sec.~\ref{sec:kelly-kelly}, we discuss a simple gambling with binary
options and introduce the notion of the Kelly criterion. 
In Sec.~\ref{sec:kelly-binary}, we study the effect of side
information on capital growth.

\subsection{Kelly criterion }\label{sec:kelly-kelly}

Let us consider simple games of gambling. 
A player has binary options to bet in a game. 
If the player wins, the casino pays twice of the betted money to the
player. 
The player repeats the game many times. 
Let us denote the fraction of money that the player bets on the $i$-th game
by $f_i$. 
The outcome of each game is assumed to be {\rm i.i.d.} in this section. 
We denote the player's capital before $i$-th game by $M_i$. 
The capital evolves as 
\begin{equation}
 M_i \longrightarrow 
\left\{
\begin{array}{l}
M_i ( 1 +  f_i) \quad {\rm win} \\
M_i ( 1 -  f_i) \quad {\rm lose} 
\end{array}
\right. .
\end{equation}
We introduce a stochastic variable $y_i \in \{1,-1\}$ 
to indicate whether the player has won or lost in the $i$-th game. 
The evolution of the capital is written as 
\begin{equation}
M_{i+1} = M_i (1+f_i y_i). 
\end{equation}
The question is: how much should the player bet in order
to make money as fast as possible? 
Kelly's recipe is the following. 
Let us define the growth rate of the player's capital during $n$ games by
\begin{equation}
 g_n(f^n) \equiv \frac{1}{n} \ln \frac{M_{n+1}}{M_1}, 
\label{eq:g-def}
\end{equation}
which is a function of the betting fractions 
$f^n \equiv \{f_1, \cdots, f_n\}$\footnote{
We use the superscript to denote the variables collectively in this
paper. 
}. 
The fractions $f^n$ are controllable variables for the player. 
Kelly's advise is to choose $f^n$ in such a way to  
maximize the average of the capital growth rate (\ref{eq:g-def}),
which is written as 
\begin{equation}
\left< g_n(f^n)  \right>_{y^n}
= 
\frac{1}{n}
\left<
\ln \frac{M_{n+1}}{M_1}
\right>_{y^n}
= 
\frac{1}{n}
\sum_{i=1}^n \left< \ln (1 + f_i y_i) \right>_{y_i}
=\frac{1}{n}\sum_{i=1}^n \left[ p_i\ln (1 + f_i)+\bar{p}_i\ln (1 - f_i) \right],
\end{equation}
where $p_i$ ($\bar{p}_i=1-p_i$) is the probability for the player to
win (lose) in the $i$-th game, 
and $\left< \cdots \right>_y$ means the average with respect to the
variable $y$. 
We denote the optimal fractions by $f^{\ast n} = \{f^\ast_1,
\cdots,f^\ast_n \}$. 
Since all the games are ${\rm i.i.d.}$, $p_1 = \cdots = p_n \equiv p$, and  
the optimal fractions are all the same, 
$f^\ast_1 = \cdots = f^\ast_n \equiv f^\ast$. 
Thus, 
$\left<g_n(f^{\ast n})\right>_{y^n}=
\left< g_1(f^\ast_1)\right>_{y_1}
\equiv\left< g(f^\ast)\right>_{y}$ and 
we have only to consider the average capital growth rate in one game. 
The solution of 
$
 d\left< g(f)\right>_y /df = 0 
$
is readily obtained as 
\begin{equation}
 f^\ast =  p - \bar p  , 
\end{equation}
where $\bar p \equiv 1 - p$. 
This choice of the betting fraction to maximize 
the average of the log of $M_{n+1} / M_1$
is called the Kelly criterion \cite{kelly1956new}.
Breiman showed that the Kelly strategy 
asymptotically surpasses other strategies in the long run
\cite{breiman1961optimal}. 
Since $f^\ast$ should satisfy $0 \le f^\ast \le 1$, 
$p$ should satisfy $p \ge 1/2 $.
$f^\ast \le 1$ holds for any value of $p$. 
When $p < 1/2$, 
the player should not bet, namely $f^\ast =0$. 
The maximum of the average capital growth rate $\langle
g(f^\ast)\rangle$ is given by 
\begin{equation}
\langle g(f^\ast) \rangle = \ln 2 - S(Y) , \label{eq:kelly-g}
\end{equation}
where $S(Y)$ is the Shannon entropy of the outcome, which can be written as
$S(Y) = S_2(p) \equiv - p \ln p - \bar p \ln \bar p$. 
If $p$ is close to $1$, $S_2(p) \simeq 0$ and 
$\langle g(f^\ast) \rangle \simeq \ln 2$, which means that the player can
almost double the capital every game. 
On the other hand, if $p \simeq 1/2$, $S_2(p) \simeq \ln 2$ and the player
can hardly increase the capital. 
Kelly noted that the capital growth rate (\ref{eq:kelly-g}) 
under the Kelly criterion
is nothing but the channel capacity of a 
binary symmetric channel with error rate $p$ (or $1-p$). 

The variance of the growth rate for general $f$ is written as 
\begin{equation}
 \begin{split}
v[g]
&= \left< g_n^2(f) \right>  - \left< g_n(f) \right>^2 \\
&=p \bar p 
\left( \ln \frac{1+f }{1- f } \right)^2 .
 \end{split}
\end{equation}
In the case of Kelly betting ($f=f^\ast$), 
\begin{equation}
 v[g(f^\ast)] = 
p \bar p 
\left( \ln \frac{p}{\bar p} \right)^2 .
\end{equation}
Although the Kelly betting is most profitable in the long run, 
it is known to be very risky in a short term and the capital suffers from large
fluctuations. In order to reduce the risk of ruin, a popular strategy is 
the ``fractional Kelly bet,''  in which a gambler bets a certain fraction
(say, $1/2$) of the Kelly fraction.

\subsection{Binary gambling with side information}\label{sec:kelly-binary}

We here consider the binary gambling in the presence of side information. 
All the games are assumed to be ${\rm i.i.d.}$ in this section, 
and we have only to consider the capital growth rate at one game. 
The result of a game is denoted by $y \in \{1,-1\}$. 
The player receives the side information $x$ before betting. 
We denote the probability distribution of the outcome $y$ for a given side
information $x$ by $P(y|x)$. 
The player determines the fraction of the capital to bet based on 
the side information, 
so the fraction depends on $x$, $f = f_x$. 
The capital evolves as 
\begin{equation}
 M_{i+1} =  M_i(1 + f_x y). 
\end{equation}
We define the capital growth rate as 
\begin{equation}
g(f_x) = \ln \frac{M_{i+1}}{M_i} = \ln (1 + f_x y).
\end{equation}

Let us calculate the maximum of the average capital growth rate in this case. 
The Kelly fraction for a fixed $x$ is given by 
$f^\ast_{x} = P(y=1|x) - P(y=-1|x)$. It follows that 
\begin{equation}
1 + f^\ast_x y = 2 P(y|x). 
\end{equation}
The average capital growth rate under the Kelly betting is given by 
\begin{equation}
\begin{split}
\left< g(f^\ast_x) \right>_{x, y} 
&= 
\sum_{x, y} P(x, y) \ln 2 P(y|x)
\\
&=
\ln 2 - S(Y) + I(X:Y), 
\end{split}
\label{eq:gbound-2}
\end{equation}
where 
$P(x,y)$ is the joint probability of $x$ and $y$, 
$S(Y) = - \left< \ln P(y) \right>$ is the Shannon entropy, 
and 
$I(X:Y)\equiv \left< \ln
\left[P(x,y)/P(x)P(y)\right]\right>$ is the mutual
information. 
This is the maximum of the average capital growth rate in the
presence of side information. 
The first two terms are the upper bound in the
absence of side information. 
The third term is the additional contribution from side information. 
With the help of side information
the upper bound of the capital growth rate is increased 
by the amount of the mutual information between 
the side information and the result of the gamble, $I(X:Y)$. 
This relation quantifies the financial value of side information. 

Let us see this in a concrete example. 
We consider the case when 
the player receives the side information of the form 
$x \in \{1, -1\}$ from an informant before betting.  
The value $1$ ($-1$) means that ``the player is going to win (lose).'' 
Unfortunately, the information from the informant can be wrong. 
Let us model the conditional probabilities as 
\begin{equation}
\begin{array}{l}
P(x=1|y=1) = P(x=-1|y=-1) = q , \\
P(x=-1|y=1) = P(x=1|y=-1) = 1-q \equiv \bar q ,
\end{array}
\end{equation}
where $q$ is the parameter to quantify the correctness of the side
information. 
The probability for a player to win (lose) is denoted by $p$ ($\bar
p$). 
The average of $g(f_x)$ is (see Table \ref{tab:prob-wire})
\begin{equation}
\left< g(f_x)\right>_{x,y}
= \sum_{x,y}
P(x,y) \ln (1 + f_x y)
= 
\bar p q \ln ( 1 - f_{-1} )
+ p q \ln ( 1 + f_1 )
+ p \bar q \ln ( 1 + f_{-1} )
+ \bar p \bar q \ln ( 1 - f_1 ).
\label{eq:gf-side-info}
\end{equation}
\begin{table}
\begin{center}
\caption{
Joint distribution of the side information and the result
 of a game, $P(x,y)$, 
and the money after betting 
in the unit of the capital before betting. 
}
\begin{tabular}{ccc}
\hline
$(x,y)$   & $P(x,y)$ & ratio \\ \hline
$(-1,-1)$ &  $\bar p q$    &  $1 - f_{-1}$   \\ \hline
$(1,1)$ & $pq$  & $1+f_1$    \\ \hline
$(-1,1)$ &  $p \bar q$ & $1+f_{-1}$  \\ \hline
$(1,-1)$ &  $\bar p \bar q$   & $1- f_1$  \\ \hline
\hline
 \end{tabular}
\label{tab:prob-wire}
\end{center}
\end{table}
We can determine the 
the fractions $f^\ast_{-1}$ and $f^\ast_{1}$ which give the maximum of
$\left< g(f_x) \right>_{x,y}$ 
by solving $\p \left< g(f_x) \right>_{x,y}/ \p f_1 = 0$ and 
$\p \left< g(f_x) \right>_{x,y}/ \p f_{-1} = 0$. The solution is 
\begin{equation}
f^\ast_{-1} = \frac{p \bar q - \bar p q }{ p \bar q + \bar p q } , \quad 
f^\ast_1 = \frac{p q - \bar p \bar q }{ p q + \bar p \bar q }. 
\end{equation}
The maximum of the average capital growth rate reads
\begin{equation}
\begin{split}
\left< g(f^\ast_x) \right>_{x,y} &= 
\bar p q \ln \frac{2 \bar p q }{p \bar q + \bar p q}
+ p \bar q \ln \frac{2  p \bar q }{p \bar q + \bar p q}
+ p q \ln \frac{2  p q }{p q + \bar p  \bar q}
+ \bar p \bar q \ln  \frac{2  \bar p  \bar q }{p q + \bar p  \bar q} \\
&= \ln 2 - S(Y) + I(X:Y),  \\
\end{split}
\end{equation}
where $S(Y) =S_2(p)= - p \ln p - \bar p \ln \bar p$, 
and we used 
$P(x=1)  = pq + \bar p \bar q$ and $P(x=-1) = p\bar q + \bar p q $.

\section{Jarzynski-type equalities for independent games of gambling}\label{sec:jarzynski}

From the discussion so far, we have learned that 
the average capital growth rate in gambling 
is bounded from above, 
\begin{equation}
g(f) \le \ln 2 - S(Y), \label{eq:g-ineq}
\end{equation}
in the case of the binary gambling without side information, for example. 
This relation can be regarded as the ``second law'' in casinos. 
If one recalls that the second law of thermodynamics is derived from an 
Integral Fluctuation Theorem (IFT), 
one might wonder that there may also be a corresponding IFT-like
equation which leads to Eq.~(\ref{eq:g-ineq}). 
As is shown below, there actually exist such equalities 
for all the situations discussed in Sec.~\ref{sec:kelly}.

Below we take the outcome of gambling $y$ to be asymmetric, 
$y \in \{R, - \bar R\}$, where $R$ and $\bar R$ are positive numbers. 
We introduce the following quantity, 
\begin{equation}
 Q(y=R) = \frac{\bar R}{R+\bar R}, \quad
 Q(y=- \bar R) = \frac{ R}{R+\bar R} .
\end{equation}
$Q(y)$ can be regarded as a probability distribution, 
since $Q(y) \in (0:1)$ and $\sum_y Q(y) = 1$. 
An important property of $Q(y)$ which is used later extensively is that 
the average of $y$ over $Q(y)$ vanishes, 
\begin{equation}
\sum_y y Q(y) = 0 .
\label{eq:prop-Q}
\end{equation}
We also denote 
\begin{equation}
s^Q_y \equiv - \ln Q(y) .
\end{equation}

\subsection{Simple binary gambling}

In the gambling of simple binary betting without side information discussed in
Sec.~\ref{sec:kelly-kelly}, the following equality holds. 
\begin{theorem}
{\rm 
Let $g(f)$ be the capital growth rate with a betting fraction $f$. 
$g(f)$ satisfies 
\begin{equation}
\left<
\exp\left[g(f) + s_y - s^Q_y\right]
\right>_y = 1 ,
\label{eq:ift-sb}
\end{equation}
where 
$
s_y = - \ln P(y) 
$, 
average of which is the Shannon entropy, $\langle s_y \rangle= S(Y)$.
\footnote{
In the context of non-equilibrium physics, 
$s_y$ is called as the trajectory (stochastic) entropy
 \cite{PhysRevLett.95.040602}
} 
}
\end{theorem}

{\it Proof. }
\begin{equation}
{\rm LHS} = 
 \left<
\left(1 + f y\right)
\frac{Q(y)}{P(y)}
\right>_y
= \sum_y (1 + f y)Q(y)
= 1  ,
\end{equation}
where we used the normalization of $Q(y)$, and 
$\sum_y Q(y) y = 0$. $\blacksquare$

Equation (\ref{eq:ift-sb}) expresses the balance between the growth rate
and the (stochastic) entropy of the results. 
We can reproduce the upper bound of the average capital growth rate
under the Kelly strategy from Eq.~(\ref{eq:ift-sb}). 
Because of the convexity of the exponential function 
($\exp \left[ \left< F \right> \right] \le \left< \exp F \right> $), 
Eq.~(\ref{eq:ift-sb}) implies 
\begin{equation}
\langle g(f) \rangle_y
\le D_{\rm KL}(P(y)||Q(y)), 
\end{equation}
where $D_{\rm KL}( \cdot || \cdot)$ is the Kullback-Leibler divergence
(see Appendix \ref{sec:app-notations} for definition). 
In the case of even-money betting ($y \in \{1, -1\}$), the bound of the
growth rate is written as 
\begin{equation}
\langle g(f) \rangle_y
\le \ln 2 - S(Y), 
\end{equation}
which equals to the upper bound of the capital growth rate (\ref{eq:kelly-g})
obtained by explicitly maximizing $\left< g(f)\right>_y$.

\subsection{Binary gambling with side information}

In the gambling with binary options under side information 
discussed in Sec.~\ref{sec:kelly-binary},  
the capital growth rate satisfies the following equality. 
\begin{theorem}
{\rm 
Let the betting fraction $f_x$, which is a function of the 
information $x$ from an informant. 
The capital growth rate $g(f_x)$ obeys
\begin{equation}
\left<
\exp\left[
g(f_x) + s_y - i_{xy} - s^Q_y
\right]
\right>_{x,y} = 1 ,
\label{eq:ift-side-info}
\end{equation}
where 
$
i_{xy} \equiv \ln 
\left[
P(x,y)/P(x) P(y)\right]
$, 
which gives the mutual information when averaged, 
$\left<i_{xy}\right> = I(X:Y)$. \footnote{
In the discussion of non-equilibrium equalities, 
$i_{xy}$ is introduced in 
Refs.~\cite{sagawa2010generalized, horowitz2010nonequilibrium}. 
Equation~(\ref{eq:ift-side-info}) is a counterpart of 
the generalized Jarzynski equality under feedback controls. 
}
}
\end{theorem}

{\it Proof. }
\begin{equation}
\begin{split}
 \left<
\exp\left[ g(f_x) + s_y - i_{xy} - s^Q_y
\right]
\right>_{x,y}
& = 
 \left<
\left(1+f_x y\right)
\frac{Q(y)}{P(y|x)}
\right>_{x,y} \\
& = 
\sum_{x,y} 
\left(1+f_x y \right)
\frac{P(x,y)Q(y)}{P(y|x)} \\
& = 
\sum_{x,y} 
\left(1+f_x y \right)
P(x)Q(y) \\
&= 1, 
\end{split}
\end{equation}
where we used the normalization of $Q(y)$, and $\sum_y Q(y) y =0$ 
in the last line. $\blacksquare$ 

Equation (\ref{eq:ift-side-info}) constrains the statistics of 
the capital growth rate, the entropy of the results and the obtained side
information.  
By using  Jensen's inequality 
$\exp \left[ \left< F \right> \right] \le \left< \exp F \right>$, 
one can derive the upper bound of $\langle g(f_x) \rangle_{x,y}$, 
\begin{equation}
\langle g(f_x) \rangle_{x,y}
\le 
D_{\rm KL}(P(y) || Q(y)) + I(X:Y). 
\end{equation}
In the case of even-money betting ($y \in \{1, -1\}$), 
\begin{equation}
\langle g(f_x) \rangle_{x,y}
\le 
\ln 2 - S(Y)
+ I(X:Y), 
\end{equation}
which coincides with Eq.~($\ref{eq:gbound-2}$). 
Again, we were able to derive the Kelly bound from the equality
(\ref{eq:ift-side-info}). 

In the presence of side information, the RHS of Eq.~(\ref{eq:ift-sb}) 
deviates from unity. Let us denote this quantity as $\gamma$, 
\begin{equation}
\gamma \equiv 
\left<
\exp\left[
g(f_x) + s_y - s^Q_y
\right]
\right>_{x,y},
\label{eq:def-efficacy}
\end{equation}
which can be also written as 
\begin{equation}
\gamma = 
1 + \sum_{x,y} f_x y P(x|y)Q(y)  .
\end{equation}
We refer to $\gamma$ as {\it efficacy} \cite{sagawa2010generalized}, 
since 
$\gamma$ is a measure of how effectively the player uses the side
information. 
This meaning is evident if one expresses $\gamma$ as 
\begin{equation}
\gamma^{-1} = 
\frac{\left< \exp\left[A_{xy} - i_{xy} \right] \right>} 
{
\left< \exp\left[A_{xy} \right]\right>
\left< \exp\left[- i_{xy}\right]
\right> 
}, 
\end{equation}
where 
$A_{xy} \equiv g(f_x) + s_y - s_y^Q$, 
and we used Eq.~(\ref{eq:ift-side-info}) and  
a trivial identity $\left< \exp\left[- i_{xy}\right] \right> = 1$. 
Thus, $\gamma$ is a measure of correlation between $A_{xy}$ and the
mutual information $i_{xy}$. 
If the side information has no effect on the capital growth, 
$\gamma=1$. 
It can be larger than $1$ when the side information contributes to
increase the capital growth. 
To see this explicitly, let us model the conditional
probability $P(x|y)$ as 
\begin{eqnarray}
 P(x=-\bar R|y=-\bar R) = P(x=R|y=R) &=& q , \\
P(x=-\bar R|y=R) = P(x=R|y=-\bar R)  &=& \bar q.
\end{eqnarray}
The parameter $q$ quantifies the correlation between $x$ and $y$. 
The value $q=1/2$ corresponds to no correlation, 
and $q=1$ means a perfect correlation. 
With parameter $q$, the efficacy is expressed as 
\begin{equation}
\gamma = 1 + 
\frac{R \bar R}{R + \bar R}
\left(
f_1 - f_{-1}
\right)\left(q- \bar q\right) .
\end{equation}
In the case of no correlation ($q = \bar q = 1/2$), 
the efficacy is unity, $\gamma=1$. 
Even if $q \neq 1/2$, 
if the player just ignores the obtained side information ($f_1 = f_{-1}$), 
$\gamma$ is again unity. 
If $q \neq  1/2$ and the player bets well,  $\gamma$ becomes larger than $1$.

As we have seen above, the efficacy depends on the choice of betting
strategy, as well as the nature of side information. 
When the gambler chooses the Kelly strategy, 
the efficacy takes a simple form, 
\begin{equation}
\gamma^\ast = 
\left<
\exp\left[i_{xy}\right]
\right>_{x,y}
=  
\sum_{x,y} P(x|y) P(y|x), 
\label{eq:efficacy-kelly}
\end{equation}
where $\gamma^\ast$ is the efficacy under the Kelly strategy. 
The upper bound of $\gamma^\ast$ can be found as 
\begin{equation}
\gamma^\ast
= 
\left<
\exp\left[i_{xy}\right]
\right>_{x,y}
= 
\left<
\exp\left[s_y - s_{y|x}\right]
\right>_{x,y}
\le
\left< 
\exp\left[ s_y \right]
\right>_{y}
= \sum_y 1
= N_Y,
\end{equation}
where $s_{y|x} \equiv - \ln P(y|x)$ and 
$N_Y=2$ is the number of possible outcomes, 
and we used the property $s_{y|x} \ge 0$. 

The efficacy can be used to detect the use of side information in gambling. 
Suppose there is a gambler in a casino and we do not know whether the gambler is
secretly using insider information. 
If we know how much the gambler bets, the results of the gamble, 
and the probability distribution of the results, 
we can calculate the quantity inside $\left< \cdots \right>$ in
Eq.~(\ref{eq:def-efficacy}). 
Observing the gambler's behavior in a number of games and taking the
average,  we can evaluate the efficacy. 
The efficacy deviates from unity if the gambler is using the side
information.

\subsection{Remarks}\label{sec:ift-remarks}

Several comments on the equalities derived in this section are in order. 
\begin{itemize}
\item 
The obtained relations hold for any choice of
      betting strategy $f$, and it is not restricted to the Kelly
      betting. 
This is similar to the case of the Jarzynski equality, 
which holds for any far-from-equilibrium  processes 
\cite{jarzynski1997nonequilibrium}. 

 \item 
In all the cases discussed above (except for the one involving the efficacy), 
the obtained equalities can be rewritten in the following form, 
\begin{equation}
\left<
e^{g(f) - g(f^\ast)}
\right>
= 1, 
\label{eq:ift-sb-2}
\end{equation}
which can be verified by noting 
$
1+f^\ast y = P(y)/Q(y)
$ 
for the case without side information or 
$
 1+f^\ast_x y = P(y|x)/Q(y)
$
in the presence of side information.
Equation~(\ref{eq:ift-sb-2}) was 
derived in the case of stock investing by Bell and 
       Cover~\cite{doi:10.1287/moor.5.2.161} in a different way. 
Since gambling can be treated as a certain form of stock investing, 
Eq~(\ref{eq:ift-sb-2}) is a special case of their result. 
In Sec.~\ref{sec:jarzynski-mem}, we generalize this equality to
       gambling with memory effects. 

\end{itemize}

\section{Jarzynski-type equalities for gambling with memory effects }\label{sec:jarzynski-mem}

So far we have assumed that all the games are independent and identical. 
Here let us consider more general cases where each gambling can 
depend on the results in the past. 
Namely, we work on the gamble with memory effects. 
We will find that 
the Jarzynski-type equalities can be generalized to those cases. 
In Sec.~\ref{sec:sb-mem}, we discuss binary gambling without side
information, and we consider the case with side information in
Sec.~\ref{sec:binary-side-mem}. 
In Sec.~\ref{sec:horse}, 
we generalize 
the equality to more general class of gambling with
arbitrary number of options and arbitrary payment functions. 
In Sec.~\ref{sec:unified}, we give a unified expression of the
equalities derived in this paper.

\subsection{Simple binary gambling}\label{sec:sb-mem}

Here we generalize the simple binary gambling without side information
to the case with memory effects. 
We define the capital growth rate during $n$ games as 
\begin{equation}
 g_n (f^n) 
\equiv 
\frac{1}{n} \ln \frac{M_{n+1}}{M_1}
= 
\frac{1}{n} \sum_{i=1}^n \ln \frac{M_{i+1}}{M_i}
= \frac{1}{n} \sum_{i=1}^n \ln \left[ 1 + f_i(y^{i-1}) y_i \right] , 
\end{equation}
where $f_i$ is the bet fraction on $i$-th game, and $y_i \in \{ R, -\bar
R\}$. 
The dependence of $f$ on $i$ means that 
the player can change the betting strategy adaptively. 
The fraction $f_i =f_i(y^{i-1})$ is a function of the results in the past,  
$y^{i-1} = \{y_1, \cdots, y_{i-1}\}$. 
The average of the capital growth rate is written as 
\begin{equation}
\left<  g_n (f^n)  \right>_{y^n}
= \sum_{y^n} P(y^n )  g_n (f^n) , 
\end{equation}
where $P(y^n) = P(y_1, \cdots, y_n)$ is the joint probability of the
results of the games. 

We define a probability distribution $Q(y^n)$ by 
\begin{equation}
Q(y^n) \equiv \prod_{i} Q(y_i), 
\end{equation}
where $Q(y_i)$ is defined by $Q(y_i=R)= \bar R / (R+\bar R)$ and 
$Q(y_i= - \bar R)=  R / (R+\bar R)$. 
Later we use the properties of $Q(y_i)$, 
\begin{equation}
 \sum_{y_i} Q(y_i) = 1,
\label{eq:qi-normalization}
\end{equation}
and
\begin{equation}
 \sum_{y_i} y_i Q(y_i) = 0.
\label{eq:qi-average}
\end{equation}
We also denote 
\begin{equation}
s^Q_{y^n} \equiv -\ln Q(y^n) . 
\end{equation}

\begin{theorem}
{\rm 
For binary gambling with memory effects, 
the capital growth rate $g_n(f^n)$ satisfies 
\begin{equation}
\left< 
\exp\left[ n g_n(f^n) + s_{y^n} - s^Q_{y^n}
\right]
\right>_{y^n} = 1, 
\label{eq:ift-mem-sb}
\end{equation}
where $s_{y^n} \equiv - \ln P(y^n) = - \ln P(y_1, \cdots, y_n)$. 
}
\end{theorem}
{\it Proof. }
\begin{equation}
\begin{split}
\left< 
\exp\left[ n g_n(f^n) + s_{y^n} - s^Q_{y^n}\right]
\right>_{y^n}
&=\sum_{y^n} \prod_{i=1}^n (1+f_i(y^{i-1})y_i )Q(y_i)  \\
&=\sum_{y^n}\left(1 + f_n(y^{n-1}) y_n \right)Q(y_n) \prod_{i=1}^{n-1} 
(1+f_i(y^{i-1})y_i )Q(y_i) .  \\
\end{split}
\end{equation}
Expanding the first bracket, the second term vanishes on summation over
$y_n$ due to Eq.~(\ref{eq:qi-average}). 
Repeating this procedure, 
\begin{equation}
\left< 
\exp\left[ n g_n(f^n) + s_{y^n} - s^Q_{y^n}\right]
\right>_{y^n}
= \sum_{y^n} Q(y^n) = 1, 
\end{equation}
where the normalization of $Q(y^n)$ [Eq.~(\ref{eq:qi-normalization})] is used. 
$\blacksquare$

Applying Jensen's inequality to Eq.~(\ref{eq:ift-mem-sb}) leads to 
\begin{equation}
\left<
g_n (f^n)
\right>_{y^n}
\le
\frac{1}{n} D_{\rm KL}(P(y^n) || Q(y^n)) . 
\label{eq:bound-with-mem}
\end{equation}
The inequality (\ref{eq:bound-with-mem}) is saturated by choosing the betting fraction as 
\begin{equation}
f_i^\ast = \frac{R P(y_i=R | y^{i-1}) -\bar R P(y_i=-\bar R | y^{i-1})}{R
 \bar R}. 
\end{equation}
With this choice of fraction,
\begin{equation}
1 + f_i^\ast y_i = \frac{ P(y_i | y^{i-1})}{Q(y_i)},  
\end{equation}
and the average capital growth rate is indeed
\begin{equation}
\begin{split}
\left<
g_n (f^{\ast n})
\right>_{y^n}
&= 
\frac{1}{n} \sum_{y^n}P(y^n) \sum_{i=1}^{n} 
\ln \frac{P(y_{i}|y^{i-1})}{Q(y_{i})}   \\
&= 
\frac{1}{n} D_{\rm KL}(P(y^n) || Q(y^n)), 
\end{split}
\end{equation}
where we used the decomposition of the joint probability 
$P(y^n) = \prod_i P(y_i | y^{i-1})$. 
In the case of even-money betting ($y_i \in \{1, -1\}$), 
\begin{equation}
\left< g_n (f^n) \right>_{y^n} 
\le \ln 2 - \frac{1}{n} S(Y^n) , 
\end{equation}
where $S(Y^n) \equiv - \left<\ln P(y^n) \right>_{y^n}$ is the entropy of
the outcomes. 
The maximum of the average capital growth rate is determined by the
entropy of $y^n$, $S(Y^n)$. 
Namely, the amount of uncertainty rules the amount of money a gambler
can make. 
The more predictable the sequence of $y_i$ is, 
the more rapidly the player's capital grows. 

{\bf Example (\underline{Markovian coin tossing})}
Let us consider a gamble using a coin. 
To begin with, the dealer places a coin on a table. 
Then, the dealer slaps the table. 
The coin flips with some probability depending on how strong the dealer 
hits the table. 
If the coin is a head (tail), the player wins (loses). 
The next game is done by hitting the table again. 
In this gamble, the results have a Markovian memory effect,
$P(y_{i}|y^{i-1}) = P(y_{i}|y_{i-1})$. 
We parametrize the dependence of the result of $i$-th game on the
previous result as  
\begin{equation}
P(y_{i}=1|y_{i-1}=-1) =  
P(y_{i}=-1|y_{i-1}=1) =\epsilon , 
\end{equation}
\begin{equation}
P(y_{i}=1|y_{i-1}=1) =  
P(y_{i}=-1|y_{i-1}=-1) = \bar \epsilon , 
\end{equation}
where $0 < \epsilon < 1$ is a constant and $\bar \epsilon \equiv 1 -
\epsilon$. 
The parameter $\epsilon$ is the flipping probability of the coin. 
Let us assume that initially the face of the coin is random, 
$P(y_0=1)=P(y_0=-1)=1/2$, 
where $y_0$ is introduced as a dummy variable and note that actual betting starts
from $n=1$. 
In this model, $P(y_i=1) = P(y_i=-1)=1/2$ 
for any $i$ and 
without using the correlations, it is 
impossible to increase the capital. 
A gambler can make money by 
exploiting the correlation of the result in the next game 
with those in the past games 
enables a gambler to make money. 
The entropy of the joint distribution 
$P(y^n) = P(y_1, \cdots, y_n)$ 
can be calculated as 
\begin{equation}
\begin{split}
S(Y^n) &= - \left< \ln P(y^n)\right>_{y^n} \\
&= 
-\sum_{i=1}^{n} \left< \ln P(y_{i}|y_{i-1})\right>_{y_i, y_{i-1}}
\\
&= 
(n-1) S_2(\epsilon) + \ln 2, 
\end{split}
\label{eq:bound}
\end{equation}
where $S_2(\epsilon) \equiv - \epsilon \ln \epsilon - \bar \epsilon \ln
\bar \epsilon$.
Note that we use the notation in which $P(y_i|y_j) = P(y_i)$ when $y_i$
is empty. 
For example, $P(y_1|y_0) = P(y_1)$ in Eq.~(\ref{eq:bound}).
The RHS of Eq.~(\ref{eq:bound-with-mem}) is given by 
\begin{equation}
{\rm RHS} = 
\frac{n-1}n\left[\ln 2 - S_2(\epsilon)\right].  
\end{equation}

On the other hand, we can explicitly maximize the 
growth rate by choosing the fraction so that 
$1+f^\ast_i y_i = 2 P(y_{i}|y_{i-1})$ is satisfied. 
The average capital growth rate with this choice of fraction is 
\begin{equation}
\begin{split}
\left< g_n(f^\ast) \right>_{y^n} 
&= \frac{1}{n} \sum_{i=1}^{n} \left< \ln (1 + f^\ast_i y_i )\right> \\
&= \frac{1}{n} \sum_{i=1}^{n} \left< \ln 2 P(y_{i}| y_{i-1}) \right>
\\
&=
\frac{n-1}n
\left[
\ln 2 - S_2(\epsilon)
\right]
, 
\end{split}
\end{equation}
which coincides with the bound (\ref{eq:bound}), and the inequality
(\ref{eq:bound-with-mem}) is saturated. 
Since the achieved growth rate (\ref{eq:bound}) is positive, 
the player has a chance to increase the capital, 
unless the flipping of the coin is completely random ({\rm i.e.}, $\epsilon=1/2$). 

The Markovian coin tossing discussed here can be mapped to the 1D
Ising model, as described in Appendix \ref{sec:app-ising}. 
The sequence of the results of the coin toss is 
identified to the configuration of Ising spins in one dimension.

\subsection{Binary gambling with side information}\label{sec:binary-side-mem}

We here work on the binary gambling with memory effects and 
side information. 
On the $i$-th game, 
the player determines the betting fraction based on 
the outcomes in the past $y^{i-1}$ and the side information $x^i = \{x_1,\cdots, x_i\}$, 
so the fraction is written as $f_i = f_i(x^i, y^{i-1})$. 
The capital growth rate during $n$ bets is defined as 
\begin{equation}
\left< g_n(f^n) \right>_{x^n, y^n} 
\equiv 
\frac{1}{n}
\left<
\ln \frac{M_{n+1}}{M_1} 
\right>_{x^n, y^n} .
\end{equation}
We can show a Jarzynski-type relation in this case as well. 
\begin{theorem}
{\rm
For binary betting with memory effects and side information, 
\begin{equation}
 \left<
\exp \left[ n g_n(f^n)  
+ s_{y^n} - i_{x^n \rightarrow y^n}
- s^Q_{y^n}
\right]
\right>_{x^n, y^n} 
= 1 , 
\label{eq:ift-mem-info}
\end{equation}
where 
\begin{equation}
s_{y^n}\equiv - \ln P(y^n), \quad
i_{x^n \rightarrow y^n}
\equiv \ln \frac{P(y^n || x^n)}{P(y^n)}, \quad 
s^Q_{y^n}\equiv - \ln Q(y^n) .
\end{equation}
The expression $P(y^n || x^n) \equiv \prod_i P(y_i|y^{i-1}, x^i)$ is 
the probability distribution 
of $y^n$ causally conditioned on $x^n$. 
}
\end{theorem}

{\it Proof. }
The LHS of Eq.~(\ref{eq:ift-mem-info}) is calculated as 
\begin{equation}
\begin{split}
 \left<
\exp \left[ n g_n(f^n)  
+ s_{y^n} - i_{x^n \rightarrow y^n}
- s^Q_{y^n}
\right]
\right>_{x^n, y^n}
&= 
\left<
\frac{
\prod_i (1 + f_i(x^i, y^{i-1}) y_i) Q(y_i)
}{P(y^n || x^n )}
\right>_{x^n, y^n}
 \\
&= 
\sum_{y^n, x^n}
\frac{P(y^n, x^n)}
{P(y^n || x^n )}
\prod_i (1 + f_i(x^i, y^{i-1}) y_i)Q(y_i)
 \\
&=
\sum_{y^n, x^n}
P(x^n || y^{n-1})
\prod_i (1 + f_i( x^i, y^{i-1}) y_i)Q(y_i)
 \\
&=
\sum_{y^n, x^n}
\prod_i (1 + f_i( x^i, y^{i-1}) y_i) P(x_i | x^{i-1}, y^{i-1} )Q(y_i)
\\
&\equiv \spadesuit , 
\end{split}
\end{equation}
where we have used the decomposition of the joint probability 
$P(y^n, x^n) = P(y^n || x^n )P(x^n || y^{n-1})$, 
and the definition of $P(x^n || y^{n-1})$ 
(see Appendix \ref{sec:app-notations}). 
Let us define 
$
A_i(y^i, x^i) \equiv (1 + f_i( x^i, y^{i-1}) y_i) Q(y_i)
P(x_i |x^{i-1},y^{i-1} )
$. 
The expression $\spadesuit$ can be written as 
\begin{equation}
\begin{split}
\spadesuit & = 
\sum_{y^n, x^n} 
\prod_{i=1}^n A_i(y^i, x^i) 
\\
&= 
\sum_{y^n, x^n}
A_n(y^n, x^n) 
\prod_{i=1}^{n-1} A_i(y^i, x^i)  \\
&= 
\sum_{y^n, x^n}
\left( 1 + f_i(x^n, y^{n-1}) y_n \right) Q(y_n) P(x_n |x^{n-1},y^{n-1} )
\prod_{i=1}^{n-1} A_i(y^i, x^i). 
\end{split}
\end{equation}
If we expand the bracket in the last line, 
the second term vanishes because it is linear
in $y_n$ and $\sum_{y_n} y_n Q(y_n)=0$. Performing the summation over $x_n$ 
(note that $\sum_{x_n} P(x_n | x^{n-1}, y^{n-1})=1$), and
repeating the same procedure for $n$ times, 
\begin{equation}
\spadesuit = 
\sum_{y^{n}, x^{n-1}} 
Q(y_n)
\prod_{i=1}^{n-1} A_i(y^i, x^i) 
= \cdots = \sum_{y^n} Q(y^n) = 1 .  \quad \blacksquare
\end{equation}

Using Jensen's inequality to Eq.~(\ref{eq:ift-mem-info}), 
we obtain the bound for the average capital growth rate as 
\begin{equation}
\left< g_n(f^n) 
\right>_{x^n, y^n}
\le 
 \frac{1}{n} D_{\rm KL}(P(y^n)||Q(y^n)) + \frac{1}{n} I_{\rm dr}(X^n \rightarrow
Y^n), 
\label{eq:bound-mem-info}
\end{equation}
where 
\begin{equation}
I_{\rm dr}(X^n \rightarrow Y^n)
\equiv 
\left<
i_{x^n \rightarrow y^n}
\right>_{x^n, y^n}
= 
\left< \ln \frac{P(y^n || x^n)}{P(y^n)} \right>_{x^n, y^n} .
\label{eq:te-def}
\end{equation}
The quantity $I_{\rm dr}(X^n \rightarrow Y^n)$ is the directed
information from $X^n$ to $Y^n$, which is a measure of
causal correlations \cite{massey1990causality, kramer1998directed, massey2005conservation}. 
Correlational measures such as the mutual information and cross
correlations between $X$ and $Y$ are symmetric under the exchange of $X$
and $Y$, and can not capture the directionality of influences. 
The directed information quantifies ``directed'' flow of information 
and 
is useful in uncovering the causal influences 
among interacting systems. 
The inequality (\ref{eq:bound-mem-info}) is saturated when the player
bets the fraction 
\begin{equation}
f_i^\ast = \frac{R P(y_i=R | y^{i-1}, x^{i}) -\bar R P(y_i=-\bar R |
 y^{i-1},  x^{i})}{R
 \bar R}. 
\end{equation}
This choice means that ``the player should bet the Kelly fraction
based on all the available information.''
With this choice of fraction, 
\begin{equation}
1 + f_i^\ast y_i = \frac{ P(y_i | y^{i-1}, x^i)}{Q(y_i)},  
\end{equation}
and one can readily check that Eq.~(\ref{eq:bound-mem-info}) is saturated. 
In the case of symmetric betting ($y \in \{1,-1\}$), 
$D_{\rm KL}(P(y^n)||Q(y^n)) = n \ln 2 - S(Y^n)$ and 
the upper bound of the average capital growth rate is written as 
\begin{equation}
\left< g_n(f^n) 
\right>_{x^n, y^n}
\le 
\ln 2 - 
\frac{1}{n} S(Y^n)
+ \frac{1}{n} I_{\rm dr}(X^n \rightarrow
Y^n). 
\label{eq:bound-mem-info-even}
\end{equation}

{\bf Example (\underline{Markovian coin tossing with side information})}
Let us discuss an extension of the Markovian coin tossing
in the previous section. 
The dealer slaps the table and tries to flip the coin, as before. 
This time, the flipping rate on the $i$-th game is a stochastic variable. 
The player infers the flipping rate (the strength for the dealer to
slap the table) from the dealer. 
We denote the flipping rate of the coin as 
\begin{equation}
 P(y_{i}|y_{i-1}, \theta_{i}) = 
\begin{cases}
\theta_{i} & y_{i} \neq y_{i-1},  \\
\bar \theta_{i} & y_{i} = y_{i-1} , \\
\end{cases}
\end{equation}
where $\theta_i \in [0:1]$ is a stochastic variable. 
The player measures the flipping rate, and determines the betting fraction
based on the measured rate. 
We assume that the initial face of the coin is random,
$P(y_0=1)=P(y_0=-1)=1/2$. 

Let us calculate the RHS of Eq.~(\ref{eq:bound-mem-info-even}). The
directed information from $\Theta^n$ to $Y^n$ reads 
\begin{equation}
\begin{split}
I_{\rm dr}(\Theta^n \rightarrow Y^n )
&= 
\left< 
\ln \frac{P(y^n||\theta^n)}{P(y^n)}
\right>_{\theta^n, y^n}
\\
&= 
\left<
\ln P(y_1|\theta_1)
\right>_{y_1, \theta_1}
+
\sum_{i=2}^n
\left< 
\ln P(y_i|y_{i-1}, \theta_i)
\right>_{y_i, y_{i-1}, \theta_i}
+
S(Y^n)
\\
&= 
- \ln 2 
- 
\sum_{i=2}^n 
\left< 
S_2(\theta_i)
\right>_{\theta_i}
+
S(Y^n), 
\end{split}
\end{equation}
where $S_2(p) \equiv - p \ln p - \bar p \ln \bar p $ is the binary
entropy function, 
and we used $P(y_i)=1/2$ since the initial state of the coin is randomly
chosen. 
If the face of the coin is independent of the measured flipping rate, 
$P(y_i|y_{i-1},\theta_i) =P(y_i|y_{i-1})$, 
then 
the directed information vanishes, 
$I_{\rm dr}(\Theta^n \rightarrow Y^n )=0$. 
Thus, the RHS of Eq.~(\ref{eq:bound-mem-info-even}) is written as 
\begin{equation}
{\rm RHS} =
\frac{n-1}{n}
 \ln 2 - \frac{1}n 
\sum_{i=2}^n
\left< 
S_2(\theta_i)
\right>_{\theta_i}. 
\label{eq:bound-rhs}
\end{equation}

The capital growth rate under the Kelly betting is calculated as follows. 
Noting that 
\begin{equation}
1 + f(y^{i-1}, \theta^i) y_i 
= 
1 + f(y_{i-1}, \theta_i) y_i 
= 2 P(y_i|y_{i-1}, \theta_i), 
\end{equation}
the average capital growth rate is 
\begin{equation}
\begin{split}
\left<
g(f^{\ast n})
\right>_{\theta^n,y^n} 
&= 
\frac 1 n
\sum_{i=1}^{n}
\left<
\ln \left( 1 + f(y^{i-1}, \theta^i) y_i \right)
\right>
\\
&= 
\frac 1 n
\sum_{i=2}^n
\left<
\ln 2 P(y_i|y_{i-1}, \theta_i)
\right>
+ \frac{1}{n} 
\left< \ln 2 P(y_1|\theta_1) \right>
\\
&= 
\frac{n-1}n
 \ln 2 
- 
\frac 1 n 
\sum_{i=2}^n
\left< S_2(\theta_i)\right>_{\theta_i}, 
\\
\end{split} 
\end{equation}
where we again used $P(y_i)=1/2$.
This coincides with Eq.~(\ref{eq:bound-rhs}) and the inequality
(\ref{eq:bound-mem-info-even}) is
saturated.

Suppose $\theta_i$ obeys a Gaussian distribution with mean $p$ and 
variance $\sigma^2$, and $\theta_i$ for different $i$  are independent. 
When $\sigma$ is small compared to  $p$ and $\bar{p}=1-p$,
the directed information $I_{\rm dr} (\Theta^n \rightarrow Y^n )$ is
approximated as 
\begin{equation}
I_{\rm dr} (\Theta^n \rightarrow Y^n ) 
\simeq \frac{(n-1) \sigma^2}{2 p \bar p}. 
\label{eq:idr-mct}
\end{equation}
The additional contribution to maximal capital growth rate 
due to the side information, 
$
\delta
\left<g^\ast\right>
\equiv 
\left<
g(f^{\ast n})
\right>_{\theta^n,y^n} 
 - 
\left<
g(f^{\ast n})
\right>_{y^n} 
$, is thus written as 
\begin{equation}
\delta
\left<g^\ast\right>
= 
\frac{1}{n} I_{\rm dr} (\Theta^n \rightarrow Y^n ) 
\simeq
\frac{n-1}{n}\frac{\sigma^2}{2 p \bar p}. 
\end{equation} 
Thus, the randomized flipping rate $\theta_i$ always helps the gambler, 
and it is an increasing function of the 
variance of the flipping rate. 

Let us find the expression for the efficacy assuming that the gambler
obeys the Kelly strategy. 
The efficacy can be defined in an analogous way in to the memoryless
case [see Eq.~(\ref{eq:efficacy-mem})]. 
As will be shown in the next section, 
the efficacy under the Kelly strategy is written as 
$\gamma^\ast = 
\left<\exp\left[i_{\theta^n \rightarrow
y^n}\right]\right>^{1/n}$. 
We can calculate $\gamma^\ast$ as 
\begin{equation}
\begin{split}
\left(\gamma^\ast\right)^n &= 
\left<\exp\left[i_{\theta^n \rightarrow
y^n}\right]\right>
= 
\left<\frac{P(y^n||\theta^n)}{P(y^n)}\right>_{y^n,x^n}
= 
\sum_{y^n,\theta^n}
\frac{P(y^n, \theta^n) P(y^n||\theta^n)}{P(y^n)}
\\
&= 
\sum_{y^n,\theta^n}
\frac{P(y_1, \theta_1) P(y_1| \theta_1)}{P(y_1)}
\prod_{i=2}^n
\frac{P(y_i, \theta_i | y_{i-1}, \theta_{i-1}) 
P(y_i|y_{i-1}, \theta_i)
}{P(y_i|y_{i-1})}  . 
\end{split}
\label{eq:gamma-star-cal}
\end{equation}
Noting that 
$
P(y_i, \theta_i | y_{i-1}, \theta_{i-1}) 
= P(y_i, \theta_i | y_{i-1}) 
$
and 
$
P(y_i, \theta_i | y_{i-1}) 
= 
P(y_i | y_{i-1}, \theta_i) 
P( \theta_i| y_{i-1}) 
= 
P(y_i | y_{i-1}, \theta_i) 
P( \theta_i) 
$, 
\begin{equation}
\left(\gamma^\ast\right)^n = 
\sum_{y^n,\theta^n}
\frac{P(y_1, \theta_1) P(y_1| \theta_1)}{P(y_1)}
\prod_{i=2}^n
\frac{P(y_i|y_{i-1}, \theta_i)^2 P( \theta_i) 
}{P(y_i|y_{i-1})} .
\label{eq:gamma-sum}
\end{equation}
The summation in Eq. (\ref{eq:gamma-sum}) can be calculated in a similar
way to the transfer matrix method for the 1D Ising model. 
In the current model, 
the marginal flipping rate that appears in Eq.~(\ref{eq:gamma-star-cal})
is written as 
\begin{equation}
P(y_i|y_{i-1}) 
= 
\sum_{\theta_i}
P(y_i, \theta_i|y_{i-1}) 
= 
\sum_{\theta_i}
P(y_i|y_{i-1}, \theta_i) P(\theta_i) 
= 
\begin{cases}
p & y_{i} \neq y_{i-1},  \\
\bar p & y_{i} = y_{i-1} .
\end{cases}
\end{equation}
Let us define matrices $B_i$ as 
\begin{equation}
[B_i]_{y_i\, y_{i-1}}  \equiv 
\sum_{\theta_i}\frac{P(y_i|y_{i-1}, \theta_i)^2 P( \theta_i) 
}{P(y_i|y_{i-1})} =
\begin{pmatrix}
\frac{\left< {\bar\theta_i}^2\right>}{ \bar p}  & \frac{\left<{\theta_i}^2\right>}{ p} \\
\frac{\left< {\theta_i}^2 \right>}{ p}      & \frac{\left<{\bar\theta_i}^2\right>}{ \bar p} 
\end{pmatrix} 
= 
\begin{pmatrix}
\frac{ {\bar p}^2 + \sigma^2 }{ \bar p}  & \frac{ { p}^2 + \sigma^2}{ p} \\
\frac{ { p}^2 + \sigma^2}{ p}      & \frac{ {\bar p}^2 + \sigma^2}{ \bar p} 
\end{pmatrix}.
\end{equation}
The matrices $B_i$ are in fact all the same and we denote them as $B
\equiv B_i$. 
Noting also that $P(y_1)= P(y_1|\theta_1) = 1/2$, 
the efficacy can be expressed by the product of those matrices as 
\begin{equation}
\begin{split}
\left(\gamma^\ast\right)^n  
&=  \sum_{a=1}^2 
\left[ \prod_{i=2}^n B_i 
\begin{pmatrix}
1/2 \\
1/2
\end{pmatrix}
\right]_{a}  \\
&=
\sum_{a=1}^2
\left[ 
\left(B\right)^{n-1}
\begin{pmatrix}
1/2 \\
1/2
\end{pmatrix}
 \right]_{a}  \\
&=
\left(
1 + \frac{\sigma^2}{p \bar p}
\right)^{n-1} .
\end{split} 
\label{eq:efficacy-calc}
\end{equation}
Thus, in the current approximation, the efficacy and the 
directed information is related by 
(from Eq.~(\ref{eq:idr-mct}) and Eq.~(\ref{eq:efficacy-calc}))
\begin{equation}
 \gamma^\ast = 
\left(
1+\frac{2}{n-1} I_{\rm dr}(\Theta^n \rightarrow Y^n)
\right)^{1 - \frac{1}n} . 
\end{equation}

\subsection{Generalization: horse races with side information}\label{sec:horse}

We here generalize the Jarzynski-type equalities to a gambling in which
the players have multiple options to bet with a help
of side information\footnote{
As another extension, we can also formulate the Jarzynski-type
equalities in gambling with more complex information structures using
the Bayesian network.  
The proof of the equality is almost the same, 
we just have to replace $P(x^n||y^n)$ with $P_c(x^n||y^n) \equiv
\prod_i^n P(x_i|{\rm pa}(x_i))$ and similarly for $f(y^n||x^n)$ and
$o(y^n||x^n)$. See Ref.~\cite{ito2013information}. 
}. 
This situation actually corresponds to horse races in which 
each result of a race can depend on the results in the past. 
Horse races can be regarded as a generalization of the cases of binary betting
discussed in the previous sections. 
For the discussion of the upper bound of the capital growth rates in
horse race, see Refs.~\cite{cover2012elements, permuter2008directed,
permuter2011interpretations}. 

Let $f(y_i)$ and $o(y_i)$ be the bet fraction and the odds on the horse
$y_i$ in the $i$-th race.  
We take $\sum_{y_i} f(y_i)=1$, 
which means that the gambler bets all one's
money in every race. 
The capital of the gambler evolves as 
\begin{equation}
 M_{i+1} = M_i f(y_i | y^{i-1}, x^i) o(y_i | y^{i-1}) , \label{eq:horseRace}
\end{equation}
where $y_i$ is the horse that won the $i$-th race, and $x_i$ is 
the side information for the $i$-th race\footnote{
The quantities $f$ and $o$ in Eq.~\eqref{eq:horseRace} correspond to 
$(1+y f)Q(y)$ and $1/Q(y)$ in the case of binary betting discussed in
the previous subsections. 
}. 
The gambler determines the betting fraction according to the past results and
the received side information, thus $f(y_i) = f(y_i | y^{i-1}, x^i)$. 
The odds $o(y_i|y^{i-1})$ are also a function of the results in 
the past. 
The gambler's capital during $n$ races is  written as 
\begin{equation}
M_{n+1} = 
M_1 \prod_{i=1}^{n} f(y_i | y^{i-1}, x^i) o(y_i | y^{i-1})  .
\end{equation}
We define the capital growth rate after $n$ races as 
\begin{equation}
g_n(f^n, o^n)
= \frac{1}{n}\ln \frac{M_{n+1}}{M_1}
= \frac{1}{n} \ln f(y^n||x^n) o(y^n), 
\end{equation}
where 
\begin{equation}
f(y^n||x^n) \equiv \prod_{i}^{n}  f(y_i | y^{i-1}, x^i) 
, \quad 
o(y^n) \equiv \prod_{i=1}^n o(y_i | y^{i-1}). 
\end{equation}

\begin{theorem}
{\rm  
In horse races with memory effects and side information, the capital growth rate $g_n$
 satisfies 
\begin{equation}
\left<
\exp \left[ n g_n(f^n, o^n) + s_{y^n} - i_{x^n \rightarrow y^n} - \ln o(y^n) \right]
\right>_{y^n, x^n}
= 1 ,
\label{eq:ift-horse}
\end{equation}
where 
\begin{equation}
s_{y^n}\equiv - \ln P(y^n), \quad
i_{x^n \rightarrow y^n}
\equiv \ln \frac{P(y^n || x^n)}{P(y^n)}. 
\end{equation}
}
\end{theorem}

{\it Proof. }
\begin{equation}
\begin{split}
\left<
\exp \left[ n g_n(f^n, o^n) + s_{y^n} - i_{x^n \rightarrow y^n} - \ln o(y^n) \right]
\right>_{y^n, x^n}  
&= 
\left<
\frac{f(y^n||x^n) }
{P(y^n || x^n)}
\right>_{y^n, x^n}   \\
&= 
\sum_{y^n, x^n}
f(y^n||x^n) 
\frac{P(y^n, x^n)}
{P(y^n || x^n)}
\\
&= 
\sum_{y^n, x^n}
f(y^n||x^n)
P(x^n||y^{n-1})
\\
&=1, 
\end{split}
\end{equation}
where we have used the decomposition of $P(x^n, y^n)$,  and 
\begin{equation}
 \sum_{y^n, x^n} f(y^n||x^n) P(x^n||y^{n-1}) = 1 . 
\end{equation}
This relation can be shown by noting $\sum_{y_i} 
f(y_i|y^{i-1})=1$ and $\sum_{x_i} P(x_i|x^{i-1},
y^{i-1})=1$. $\blacksquare$

Using Jensen's inequality, we obtain the bound for capital growth, 
\begin{equation}
\left< g_n(f^n, o^n) \right>_{y^n,x^n}
\le \frac{1}{n} \left< \ln o(y^n) \right>_{y^n, x^n} 
-\frac{1}{n} S(Y^n) + \frac{1}{n}I_{\rm dr }(X^n \rightarrow Y^n),
\label{eq:bound-horse}
\end{equation}
where we have used 
$\left< i_{x^n \rightarrow y^n} \right>_{x^n,y^n} =
I_{\rm dr}(X^n \rightarrow Y^n) $. 
Under the ``fair and uniform'' odds ($o(y_i | y^{i-1}) = M$ for any $i$,
where $M$ is the number of horses in a race), 
\begin{equation}
\left< g_n(f^n, o^n) \right>_{y^n,x^n} 
\le \ln M - \frac{1}{n} 
S(Y^n) + \frac{1}n I_{\rm dr }(X^n \rightarrow Y^n) , 
\end{equation}
which reproduces the results obtained in 
Refs.~\cite{permuter2008directed, permuter2011interpretations}. 
The upper bound (\ref{eq:bound-horse}) is achieved when the player
chooses the betting fraction as 
\begin{equation}
f(y_i|y^{i-1}, x^i)  = P(y_i|y^{i-1}, x^i). 
\end{equation}

The efficacy can also be straightforwardly extended to the case with
memory effects, 
which we define as 
\begin{equation}
\gamma 
\equiv 
\left(
\left<
\exp \left[ n g_n(f^n, o^n) + s_{y^n} - \ln o(y^n) \right]
\right>_{y^n, x^n}
\right)^{1/n}. 
\label{eq:efficacy-mem}
\end{equation}
The efficacy is also written as 
\begin{equation}
\gamma^{-n}
= 
\frac{
\left<
\exp\left[A_{x^n y^n} - i_{x^n \rightarrow y^n}\right]
\right>_{y^n, x^n}
}
{
\left< \exp\left[A_{x^n y^n}\right] \right>_{y^n, x^n}
\left< \exp\left[ - i_{x^n \rightarrow y^n}\right] \right>_{y^n, x^n}
},
\label{eq:mem-efficacy-2}
\end{equation}
where 
$
 A_{x^n y^n} \equiv  n g_n(f^n, o^n) + s_{y^n} - \ln o(y^n) 
$, and we used
$\left< \exp\left[ - i_{x^n \rightarrow y^n}\right] \right>_{y^n, x^n}=
1$ and Eq.~\eqref{eq:ift-horse}.
From the expression (\ref{eq:mem-efficacy-2}), it is clear that 
the efficacy measures correlation between the capital growth rate
and information flow. 
If the gambler is to take the Kelly strategy, the efficacy is written as 
$
\gamma^\ast
= 
\left(
\left<
\exp\left[ i_{x^n \rightarrow y^n}\right]
\right>_{y^n, x^n}
\right)^{1/n},  
$
which can be shown as 
\begin{equation}
\begin{split}
\left(\gamma^\ast\right)^n
&= 
\left<
\exp \left[ n g_n(f^n, o^n) + s_{y^n} - \ln o(y^n) \right]
\right>_{y^n, x^n}  \\
&= 
\left<
\frac{f(y^n||x^n) }{P(y^n)}
\right>_{y^n, x^n}   \\
&= 
\left<
\frac{P(y^n||x^n) }{P(y^n)}
\right>_{y^n, x^n}   \\
&= 
\left<
\exp \left[ i_{x^n \rightarrow y^n}\right]
\right>_{y^n, x^n} , 
\end{split}
\label{eq:efficacy-mem-kelly}
\end{equation}
where the fact that $f(y^n||x^n)=P(y^n||x^n)$ holds for the Kelly
betting is used. 
The upper bound of $\gamma^\ast$ is found as 
\begin{equation}
\begin{split}
\gamma^\ast
&= 
\left(
\left<
\exp\left[i_{x^n \rightarrow y^n}\right]
\right>_{y^n, x^n}
\right)^{1/n} \\
&= 
\left(
\left<
\exp\left[s_{y^n} - s_{y^n||x^n}\right]
\right>_{y^n, x^n}
\right)^{1/n} 
\\
&
\le
\left(
\left< 
\exp\left[ s_{y^n} \right]
\right>_{y^n}
\right)^{1/n} \\
&=
\left(\sum_{y^n} 1 \right)^{1/n} \\
&=  M , 
\end{split}
\end{equation}
where $s_{y^n||x^n} \equiv - \ln P(y^n||x^n)$ and 
and we used the property $s_{y^n||x^n} \ge 0$.

\subsection{Unified expression of the Jarzynski-type equality in
  gambling}\label{sec:unified}

Here we give the unified expression of the Jarzynski-equalities 
in gambling. 
The equalities 
of the form $\left< \cdots \right> = 1$ 
shown in this paper can be written in the following form, 
\begin{equation}
\left<
e^{  n \left[ g_n(f^n) - g_n(f^{\ast n})   \right]}
\right>_{y^n, x^n}
= 1 ,  
\label{eq:ift-unified}
\end{equation}
where $f^n$ are the betting fractions for $n$ games, 
and $f^{\ast n}$ are the betting fractions under the Kelly strategy. 
The betting fractions $f^n$ can depend on the results in the past, and
also on the side information. 
This equality holds regardless of the choice of the betting fractions
$f^n$. 
Equation (\ref{eq:ift-unified}) is a generalization of Bell and Cover's
result \cite{doi:10.1287/moor.5.2.161} to gambling with memory effects.

\section{Summary and outlook}\label{sec:summary}

In this paper, we derived novel Jarzynski-type equalities in gambling. 
Based on those relations, we discussed how the side information or the
memory effects affect the capital growth of a gambler. 
The main results of this paper are summarized as follows:
\begin{itemize}
\item Simple binary gambling $\rightarrow $ Eqs.~(\ref{eq:ift-sb}) and ~(\ref{eq:ift-mem-sb})
\item Binary gambling with side information
       $\rightarrow $ 
      Eqs.~(\ref{eq:ift-side-info}) and (\ref{eq:ift-mem-info})
\item Horse races with side information and memory effects
      $\rightarrow$ Eq.~(\ref{eq:ift-horse})
 \item 
Introduction of the efficacy $\rightarrow$
       Eqs.~(\ref{eq:def-efficacy}) and (\ref{eq:efficacy-mem})
\end{itemize}
The equality for horse races is the most general one. 
We also gave the unified expression of the Jarzynski-type relations
shown this paper (except for the one involving the efficacy) 
in Eq.~(\ref{eq:ift-unified}). 
Those equalities reproduce the known Kelly bounds of the capital growth
rates by applying Jensen's inequality. 
In the case of betting with side information, the financial value of the
information is quantified by the mutual information in independently
repeated gambling and by the directed information for gambling with memory
effects. 
We defined the efficacy in Eqs.~(\ref{eq:def-efficacy}) and 
(\ref{eq:efficacy-mem}), 
which is a measure of how well the gambler makes use of the side information. 
We showed that 
the efficacy allows for a simple expression 
[Eqs. (\ref{eq:efficacy-kelly}) and (\ref{eq:efficacy-mem-kelly})]
when the gambler bets under
the Kelly strategy, and discussed its upper bound. 

Finally, let us comment on possible future directions. 
\begin{itemize}
 \item {\it Application to actual gambling.} 
An interesting direction is the analysis of actual gambling. 
A good candidate would be blackjack. 
In blackjack, a player can improve the return by exploiting 
the information of the cards which are already open to the floor. 
This method is called ``card counting'' \cite{thorp1966beat}. 
The player assigns a number for each card, and sums up the number for
all the dealt cards. 
The sum is called the count, which is a measure of the card composition of the shoe. 
The true count, which is the count divided by the number of remaining
       decks, is indicative of the expected return. 
There are many possible choices of the weights on the cards. 
The performance of a counting system is evaluated by a quantity
called betting correlation. 
It is possible to estimate the growth rate of wealth under the Kelly criterion as a
function of betting correlations\footnote{
Mathematical aspects of blackjack are comprehensively analyzed in a recent book
by Werthamer \cite{werthamer2009risk}. 
}. 
\item {\it Extension to stock investing.} 
Although we focused on gambling in this paper, extension of such equalities
to stock investing is quite interesting. 
Such concepts as the Jarzynski-equality and the fluctuation
theorems developed in non-equilibrium physics might be useful in
uncovering the role of information in the financial world.
\end{itemize}

\acknowledgements

Y.~Hirono is grateful to S. Nakayama for useful discussions and careful
reading of the manuscript. 
Y.~Hirono is supported by JSPS Research Fellowships for Young Scientists. 
This work is partially supported  by the RIKEN iTHES Project.
This work is also supported by JSPS Strategic Young Researcher Overseas Visits Program for Accelerating Brain Circulation (No. R2411).


\appendix

\section{Notations and definitions }\label{sec:app-notations}

Here we summarize notations and definitions of information-theoretical
quantities used in the text. 

A realization of a stochastic variable $X$ is represented by its small
letter, $x$ in this case. 

Let $\{ x_i\}$, $\{y_i\}$ be time-sequences of stochastic variables. 
A variable with a superscript $n$ indicates variables 
from $1$ to $n$ collectively, 
\begin{equation}
 x^n \equiv \{ x_1, \cdots, x_n \}. 
\end{equation}

We assume that the dependences of the variables $x^n$ are causal, 
by which we mean that 
the probability distribution $P(x_i)$ is dependent on $x_j$ only if
$j<i$. 
The joint probability $P(x^n)$ is decomposed as 
\begin{equation}
 P(x^n) = \prod_i P(x_i | x^{i-1}) , 
\end{equation}
where $P(x|y)$ is the conditional probability. 

The average over variables $\{x,y,\cdots\}$ is expressed by $\left< \cdots
\right>_{x,y,\cdots}$. 
Subscript may be omitted, in that case the average is taken over all the
stochastic variables. 

The Shannon entropy of $X^n = \{X_1, \cdots, X_n \}$ is 
\begin{equation}
 S(X^n) \equiv - \left< \ln P(x^n) \right>_{x^n} 
=  - \sum_{x^n} P(x^n) \ln P(x^n)  .
\end{equation}

The Kullback-Leibler divergence of a distribution $Q(y)$ from another
distribution $P(y)$ is defined by
\begin{equation}
 D_{\rm KL}\left(P(y)||Q(y) \right) \equiv \sum_y P(y) \ln
  \frac{P(y)}{Q(y)}. 
\label{eq:kl-div}
\end{equation}

The mutual information between the stochastic variable $X$ and $Y$ is
defined by 
\begin{equation}
I(X:Y) \equiv 
\left< \ln \frac{P(x,y)}{P(x)P(y)} \right>_{x,y}
= \sum_{x,y} P(x,y) \ln \frac{P(x,y)}{P(x)P(y)}.
\end{equation}

We used the following causal conditioning notations developed by Kramer
\cite{kramer1998directed}. 
The probability distribution of $x^n$ causally conditioned on $y^{n-d}$
is denoted as 
\begin{equation}
P(x^n || y^{n-d}) \equiv \prod_{i=1}^n P(x_i | x^{i-1}, y^{i-d}). 
\end{equation}
We use a convention that, if $i-d \leq 0$, $y^{i-d}$ is set to null. 
Mostly, the cases with $d=0,1$ are used:
\begin{equation}
P(x^n || y^{n}) = \prod_{i=1}^n P(x_i | x^{i-1}, y^{i}), 
\end{equation}
\begin{equation}
P(x^n || y^{n-1}) = \prod_{i=1}^n P(x_i | x^{i-1}, y^{i-1}).
\end{equation}

The joint probability of $x^n$ and $y^n$ is decomposed as 
\begin{equation}
P(x^n , y^n) = P(x^n || y^n) P(y^n || x^{n-1}) .
\end{equation}
\begin{equation}
\begin{split}
\because
P(x^n , y^n) 
&= \prod_i P(x_i, y_i | x^{i-1}, y^{i-1}) \\
&= \prod_i 
P(x_i | x^{i-1}, y^i) 
P(y_i | x^{i-1}, y^{i-1}) \\
&= P(x^n || y^n) P(y^n || x^{n-1}). 
\end{split}
\end{equation}

The causally conditional entropy is defined as 
\begin{equation}
 S(X^n || Y^n) 
\equiv 
-  \left< \ln P(x^n || y^n)\right>
= \sum_{i=1}^n S(X_i | X^{i-1}, Y^i) .
\end{equation}

The directed information, introduced by Massey
\cite{massey1990causality}, is defined as 
\begin{equation}
I_{\rm dr}(Y^n \rightarrow X^n) \equiv S(X^n) - S(X^n || Y^n) .
\end{equation}
It can be explicitly written as 
\begin{equation}
I_{\rm dr}(Y^n \rightarrow X^n) = \left< \ln \frac{P(x^n ||
				   y^n)}{P(x^n)} \right>
=
\sum_i
\left<
\ln \frac{P(x_{i+1} | x^{i}, y^{i+1})}{P(x_{i+1}|x^{i})}
\right> .
\end{equation}

\section{Markovian coin tossing and 1D Ising model
}\label{sec:app-ising}

We here show the equivalence of the Markovian coin tossing discussed
in Sec.~\ref{sec:sb-mem} with the 1D Ising model. 
Without loss of generality, we can 
parametrize the conditional probability $P(y_{i+1}|y_i)$ as 
\begin{equation}
P(y_{i+1}|y_i) = \frac{\exp\left[\beta J \ y_{i+1} y_i \right] }{2 \cosh
 \beta J}  .
\end{equation}
One can see
$0 < P(y_{i+1}|y_i) < 1$ and 
the normalization condition
$\sum_{y_{i+1}}P(y_{i+1}|y_i)=1$ is satisfied. 
The new parameter $J$ can be related to the flipping rate $\epsilon$ as 
\begin{equation}
\beta J = \frac{1}{2} \ln \frac{\bar \epsilon}{\epsilon} .
\end{equation}
By rewriting the normalization condition of $P(y^n)$ in the following
way, 
the correspondence to the Ising model is evident:
\begin{equation}
\begin{split}
1 &= \sum_{y^n} P(y^n)  \\
&= \sum_{y^n} \exp \left[ \sum_i \ln P(y_{i+1}|y_i) \right]\\
&= 
\frac{1}{\left(2 \cosh \beta J\right)^n} 
\sum_{y^n} \exp \left[\sum_i \beta J \ y_{i+1}y_i \right] \\
&\equiv 
\frac{{\rm tr}\left[ e^{- \beta H} \right] }{Z} .
\end{split}
\end{equation}
Thus, the numerator is the definition of the 
partition function of the Ising model without external fields. 
The average of the exponential of $g$ is written as 
\begin{equation}
\begin{split}
\left< \exp \left[n g_n \right] \right>_{y^n}
&= \sum_{y^n} P(y^n)  \prod_i (1 + f_i y_i) 
\\ 
&=
\frac{1}{\left(2 \cosh \beta J\right)^n} 
\sum_{y^n} \exp \left[\sum_i \beta J \ y_{i+1}y_i 
+  
\sum_i \ln (1 + f_i y_i) 
\right] . 
\end{split}
\end{equation}
When $f_i(y_i|y^{i-1})$ is independent of $y^{i-1}$, 
the numerator of RHS is the partition function of 
the Ising model in a weird form of magnetic field. 
In one dimension, the symmetry breaking never occurs in the Ising
model at finite temperature. 
In the context of the Markovian coin tossing, 
the absence of symmetry breaking corresponds to the fact that, 
at finite values of $\epsilon$, the coin flips after finite number of
trials, and the ``magnetization'' always vanishes, 
\begin{equation}
\lim_{n \rightarrow \infty} \frac{1}{n}\sum_i 
\left< y_i \right> = 0 .
\end{equation}

\bibliographystyle{unsrt}
\bibliography{refs}


\end{document}